\documentclass[11pt]{amsart}
\usepackage[margin=3cm]{geometry}
\usepackage{amsaddr}
\usepackage{amsfonts,latexsym}
\usepackage{amsthm}

\usepackage[rgb]{xcolor}
\usepackage{enumitem}
\usepackage{pdfsync}
\usepackage{hyperref}
\hypersetup{
       colorlinks,
           citecolor=black,
	       filecolor=black,
	           linkcolor=black,
		       urlcolor=black
		     }
\usepackage{xcolor}
 \usepackage{mathtools} 
\usepackage{amsmath,amssymb}

\newtheorem*{theorem*}{Theorem}
\newtheorem{theorem}{Theorem}[section]

\newtheorem{lemma}[theorem]{Lemma}

\theoremstyle{definition}
\newtheorem{defn}{Definition}[section]

\theoremstyle{remark}

\newcommand{\R}{\mathbb{R}}

\newcommand{\C}{\mathbb{C}}

\newcommand{\g}{\mathbf{g}}

\newcommand{\M}{\mathcal{M}}

\newcommand{\BF}[1]{\mathbf{{#1}}}

\newcommand{\PM}[1]{\begin{bmatrix} #1 \end{bmatrix} }

\makeatletter
\newcommand{\newreptheorem}[2]{\newtheorem*{rep@#1}{\rep@title}\newenvironment{rep#1}[1]{\def\rep@title{#2 \ref*{##1}}\begin{rep@#1}}{\end{rep@#1}}}
\makeatother
\newreptheorem{theorem}{Theorem}

\title[Hydrogenic atoms on spacetimes with mild singularities]{Spectrum of the Dirac Hamiltonian for hydrogenic atoms on spacetimes with mild singularities}
\author{Moulik Kallupalam Balasubramanian}
\address{Department of Mathematics\\ Rutgers University\\ Piscataway, New Jersey, U.S.A}
\date{September 23, 2020}

\begin{document}
\maketitle
\begin{abstract}
  We model a single-electron ion (hydrogenic atom) as a static,  spherically symmetric electrovacuum spacetime in which the nucleus is treated as a timelike line-singularity and the electron is treated as a test particle following Dirac's equation. The spacetime is a solution of Einstein-Maxwell equations 
  with a non-linear vacuum law. An example is Hoffmann's spacetime
  obtained using the Born-Infeld law. The Dirac Hamiltonian is 
  shown to be essentially self-adjoint, independent of the atomic number. The essential spectrum and absolutely continuous spectrum are
  the same as in Dirac's Hamiltonian on Minkowski space with a
  Coulomb potential. Presence of 
  infinitely many eigenvalues is shown and a clustering result is obtained.\end{abstract}

\section{Introduction}
Coupling of gravity and quantum mechanics is an area of active research (
\cite{z2014zglknem}
\cite{kz2014dpezgkn}
\cite{kz2014novelqmide}
\cite{franklin2011approximate}
\cite{carley2006nonperturbative}
\cite{belgiorno2000nakedanom}
\cite{belgiorno_rn}
\cite{finsmoyau2000nonexitimeperDirRN}
\cite{cohen1982general}
).
We explore it in the context of a hydrogenic 
atom, in which the nucleus is modelled by the line-singularity of an
electrovaccum spacetime and the electron as a test particle. Most of the results here are part of the author's Ph.D. thesis (\cite{kallupalam2015scalar}).

Schr\"odinger's (or, if spin is included, Pauli's) equation with a Coulomb potential is the non-relativistic 
quantum mechanical model for the electron in a hydrogenic atom.
The corresponding special-relativistic model is Dirac's equation with a Coulomb potential. Both of these cases are well-studied,
and the spectral properties including explicit formula for eigenvalues and
eigenfunctions are known. 
In these models, however, the Coulomb field is highly singular and of infinite energy, making the coupling to 
relativistic gravity problematic. The situation is similar to the appearance of infinities in the classical electrodynamics 
of point particles. To address those difficulties, Born and Infeld (\cite{bi1934}) proposed modifying the
electromagnetic vacuum law in Maxwell's equations. The result was
a non-linear constitutive law that led to an electric potential with Coulomb-like asymptotics near infinity, 
but one which, at the nucleus, took on a finite value. 

The original Born-Infeld model was not concerned with the effect of gravity. For that, one needs to  
couple Einstein's equations to Maxwell's. Under Maxwell's original (linear) vacuum law the static, spherically-symmetric, asymptotically flat solution of this system is  the Reissner-Weyl-Nordstr\"om (RWN) spacetime, where the super-extremal (or naked) case is of primary interest
as a model for the spacetime of a particle. Both the spacetime and the
electric potential are still highly singular at the center. 
The total electric energy is still infinite.

Under similar conditions, Hoffman solved Einstein-Maxwell equations with the Born-Infeld vacuum law (equation 39 \ in  \cite{hoffmann_solution}).
The metric coefficients no longer blow up and the spacetime has a ``conical singularity'' at the center. Tahvildar-Zadeh \cite{stz2011} considered all
electromagnetic theories derivable from a Lagrangian, and identified a class of them where the central singularity of the 
corresponding spacetime is as mild as possible. This class includes the Born-Infeld Lagrangian. These spacetimes are parametrized by one 
function and two real parameters.
The function 
represents the reduced Hamiltonian of the electromagnetic theory, and should satisfy certain properties. The real parameters represent the spacetime's (particle's) mass and charge. 
The non-dimensionalized mass-to-charge ratio is required to be small. In this article, we deal with such spacetimes.

We consider the interaction with an electron, which is a spin-half particle. Ideally, one should find an electromagnetic spacetime
that has two line singularities, one corresponding to the nucleus and one to the electron.  
As a first step, we consider the electron as a ``test particle'' on a spacetime with one line  singularity that represents a fixed nucleus (in Born-Oppenheimer approximation).
The background spacetime enters the
guiding equation of the electron, because the wave function of the electron obeys Dirac's equation
set up on this curved spacetime. The effect of the electron on the spacetime is ignored.

 \subsection{Notation}
 
 We use Gaussian CGS units throughout. We denote the mass and charge of the electron by $m, -e$ (so, $e>0$) respectively, the mass and charge of the spacetime by $M$, $Q$ respectively, the
speed of light in vacuum by $c$, Newton's universal constant of gravitation by $G$, and the dimensionless 
mass-to-charge ratio of the spacetime to be 
\begin{equation}
\epsilon :=\frac{\sqrt{G} M }{|Q|}.
\label{eq:Epsi} 
\end{equation}

An ion with only one electron is called a hydrogenic atom. We 
denote by $Z$ the atomic number, $N_n$ the number of neutrons in the 
atom, $M_p$ mass of a proton, $M_n$ mass of a neutron. Then,
  $Q =  Ze$
and $  M = ZM_p + N_n M_n.$

\subsection{Main Results}

We work in the setting of Hoffman-type spacetimes described in \cite{stz2011}. 

 In theorem \ref{thm:Dirac-esa-th} we prove that the Dirac Hamiltonian defined on compactly supported functions is essentially self-adjoint. This is significant, 
 as evolution and eigenvalues depend on the self-adjoint operator, not on the formal one. The Dirac-Coulomb Hamiltonian on super-extremal RWN is not essentially self-adjoint (section III, \cite{belgiorno_rn}).  On Minkowski space with Coulomb potential, essential self-adjointness is  related to the magnitude of the 
 nuclear charge. Various regimes that arise are discussed in \cite{hogreve2012overcritical}.  In our case, apart from the smallness condition on the mass-to-charge ratio which is required for the spacetime itself to be defined, no restriction on the magnitude of the nuclear charge is necessary.

In theorem \ref{thm:Ess-Spec}, we prove that the essential spectrum is 
same as  that of the Dirac-Coulomb Hamiltonian on Minkowski.
In theorem \ref{thm:PAC-DIR}, we prove that outside of $[-mc^2,mc^2]$ the spectrum is purely absolutely continuous.
In theorem \ref{thm:EValExistence}, we prove that the discrete spectrum is non-empty and infinite. This is again independent of the magnitude of the nuclear charge. Theorem \ref{thm:EValClustering} is a result about clustering of eigenvalues.

We only prove the statements for the radial part of the full operator, reduced under spherical symmetry. Much simplicity is achieved from the fact that the electric potential and metric coefficients are finite everywhere, even at the location of the charge.

The remainder of this article is structured as follows. In section \ref{sec:pf-str}, we review the spacetime from \cite{stz2011} and arrive at the Dirac Hamiltonian reduced using spherical symmetry.
Section \ref{sec:esa} is about essential self-adjointness. Section \ref{sec:spec} is
divided into subsections on the essential spectrum,
continuous spectrum, discrete spectrum and clustering of eigenvalues. Finally, in section \ref{sec:summary}, we summarize the main results and provide an outlook.

\section{Dirac equation on generalizations of the Hoffmann solution} \label{sec:pf-str}
The main equations of general relativity are Einstein's equations. The unknowns in
Einstein's equations are a four-dimensional manifold $\M$, 
a Lorentzian metric $\g$ on $\M$, and in case of non-vacuum spacetimes,
a collection of matter fields giving rise to an  energy-momentum tensor $\BF{T}$. 
In Gaussian CGS units, the equations read 
\begin{equation}
  R_{\mu\nu} - \frac{1}{2} R g_{\mu\nu} = \frac{ 8 \pi G  }{c^4} T_{\mu\nu}. \nonumber
  \label{eq:Eins}
\end{equation}
Here, $R_{\mu\nu}, R$ are respectively the Ricci and scalar curvatures of the metric.

In \cite{stz2011}, Einstein's equations coupled to a non-linear electromagnetic
theory is solved assuming that the spacetime is static, spherically symmetric, asymptotically
Minkowski, with zero magnetic charge. The solution is required to have the 
mildest possible singularity at the center of symmetry, and that the total energy
of the spacetime is equal to the energy carried by the electric field (that is, the ADM mass of the spacetime is entirely of electromagnetic origin).

  The solution in \cite{stz2011} is a spacetime with an electric potential defined on it. It is specified by three entities:
a parameter $M>0$ with units of mass,
a parameter $Q \neq 0$ with units of charge and 
    a $C^2$ function called \emph{reduced Hamiltonian} 
$    \zeta: (0,\infty) \rightarrow (0,\infty).$
The function $\zeta$ specifies the electromagnetic theory. 
  It has to satisfy conditions that correspond to physical requirements. These are: 
     $ \zeta(\mu) = \mu + O(\mu^{5/4})  \mbox{ as } \mu \rightarrow 0$
    (in the weak field limit $\zeta$ agrees with that of linear Maxwell $\zeta_0(\mu):=\mu$),
    $  \zeta' \geq 0,  \zeta-\mu\zeta' \geq 0$ (dominant energy condition is satisfied),
    $\zeta'(\mu)+2\mu\zeta''(\mu) \geq 0$ ($\zeta$ can be derived from a single-valued Lagrangian),
    $
       I_{\zeta} $ defined by $ 2^{-\frac{11}{4}}\int_0^\infty \mu^{-7/4} \zeta(\mu) d\mu$ is finite 
       (total electric energy is finite)  and finally, 
    there exists positive constants $\mu_0,J_\zeta,K_\zeta,L_\zeta$ such that 
       $\forall \mu > \mu_0$, 
     $J_\zeta \sqrt{\mu} -K_{\zeta} \leq \zeta(\mu) \leq J_{\zeta} \sqrt{\mu}$
     and 
      $\frac{J_{\zeta}}{2\mu^{1/2}} - \frac{L_{\zeta}}{\mu} \leq \zeta'(\mu) \leq \frac{J_{\zeta}}{2 \mu^{1/2}}$
    (singularity is the mildest possible). 
    
   We note that using  Gr\"onwall's inquality,  $\zeta - \mu \zeta' \geq 0$  and  $\zeta \sim \mu$ for small $\mu$, it follows that $\zeta(\mu) \leq \mu \quad \forall \mu >0.$

   The traditional Maxwell's equations result from taking $\zeta(\mu)=\mu$, which leads to a linear system.
However, it fails to satisfy the last two  conditions above. With this $\zeta$, the solution
is the RWN spacetime, which is the maximal analytic extension of 
  \begin{equation}
    ds^2 = -a(r)c^2 \ dt^2 + a(r)^{-1} dr^2 +r^2 (d\theta^2 +\sin^2(\theta) d\phi^2) \label{eq:reis}
  \end{equation}
  where $a(r)=  ( 1- 2 \frac{GM}{c^2} \frac{1}{r} + \frac{G}{c^4} \frac{Q^2}{r^2} ) $ with an
 electric potential given by
 $   \varphi(r) = \frac{Q}{r}.$
  If $\epsilon< 1 $, then $a(r)>0  \ \forall r \in (0,\infty)$, 
  and the RWN spacetime is said to be \emph{super-extremal}, or \emph{naked}. This spacetime is highly singular 
  at $r=0$.

An important example 
that does satisfy all of the conditions is the Born-Infeld Hamiltonian,
\begin{equation}
  \zeta_{BI}(\mu):= \sqrt{1+2\mu} -1.
  \label{eq:zeta-BI}
\end{equation}

Given $M,Q,\zeta$, it is necessary to use a scaled  $\zeta$ to ensure that the energy of the electric field
is equal to that of the ADM mass $M$: 
\begin{align}
  \zeta_{\beta} &:= \frac{1}{\beta^4} \zeta(\beta^4 \mu) &   \beta &:= \frac{ (| Q|)^{3/2} I_{\zeta}}{Mc^2}.
  \label{eq:zeta-beta}
\end{align} 

Note, $\beta$ has the dimension of $\frac{Length}{ \sqrt{Charge}}$, since $\zeta$, $I_{\zeta}$ are dimensionless. Further, the non-dimensionalized mass-to-charge ratio is required to be small (with $J_\zeta$ as in the requirements above): 
   \begin{align}
     \epsilon^2 &< \frac{1}{A(\zeta)}, &   A(\zeta) &:=\frac{\sqrt{2}J_{\zeta}}{I_{\zeta}^2}.
     \label{eq:small-ness}
   \end{align}

This condition is satisfied by the hydrogen atom with the electric field given by the Born-Infeld potential.
In Gaussian CGS units, approximately, $e=4.80320\cdot 10^{-10}$ esu, $M=(1.67262+1.67493)\cdot 10^{-24}$ g,
and $G=6.67408 \cdot 10^{-8}  \mbox{cm}^3 \mbox{g}^{-1} \mbox{s}^{-2}$, and so,
for a hydrogen atom, $\epsilon = 1.80049 \cdot 10^{-18}$ and its square is of the order $10^{-36}$.
Further,  $J_{\zeta}= \lim_{\mu \rightarrow \infty } \zeta(\mu)/ \sqrt{\mu}$,
which evaluates to $\sqrt{2}$. And, $I_{\zeta}=1.23605$ approximately,
using which $1/A(\zeta)$ is $0.76391$. This is larger than $10^{-36}$.

The spacetime is diffeomorphic to $\R \times (\R^3\setminus \BF{0}) $. We use coordinates
 $(t,r,\theta, \phi) \in (-\infty, \infty) \times (0, \infty ) \times (0, \pi) \times (0, 2 \pi)$. They can be described 
 directly in terms of the spacetime in the following way.
 First, $t$ is a time function such that the hypersurface orthogonal 
Killing field is $K=\frac{\partial}{c \partial t}$, and $r$ is the 
area-radius coordinate (that is, for a point $p$ in the spacetime, $r(p)=\sqrt{\frac{\mbox{Ar}(p)}{4\pi}}$ , where $\mbox{Ar}(p)$ is equal to the area of the orbit of the rotation group
$SO(3)$ that passes through $p$) and $(\theta,\phi)$ are 
spherical coordinates on the orbit sphere. 

The line element of the metric $\BF{g}$ of the spacetime is 
\begin{equation} \label{eq:background}ds_{\BF{g}}^2= -f(r)^2c^2 dt^2 + f(r)^{-2} dr^2 + r^2 d\theta^2  + r^2 \sin^2\theta  \ d\phi^2, \end{equation} 
where, 
\begin{equation} \label{eq:xiandm} f(r)^2:= 1 - \frac{ 2G}{r c^4}\int_{0}^{r}\zeta_{\beta}\left(\frac{Q^2}{2s^4}\right)s^2 ds. \end{equation}
 The electric potential is
 \begin{equation}
   \varphi(r) =  Q \int_{r}^{\infty} \zeta'_\beta\left( \frac{Q^2}{2s^4} \right) \frac{ds}{s^2}.
   \label{eq:M-el-pot}
 \end{equation}
 Here, $|\varphi| $ is a decreasing function of $r$ and $f(r)$ is an increasing function of $r$. 
The smallness of $\epsilon$ ensures that the metric coefficient $f(r)^2 >0$.
The metric is singular at $r=0$, for any choice of $\zeta$ satisfying the conditions mentioned earlier. The singularity is \emph{conical}: $\lim_{r\rightarrow 0} f(r)$ exists but is not equal to $1$.

We now state the asymptotics from \cite{stz2011} with $G, c$ added.
Below, $A, B$ are constants depending on $\zeta$, with $A$ as defined earlier.
Near $r=0$,
\begin{eqnarray}
  \label{eq:asymptotics-near-0}
  f(r)^2 &=& (1-A\epsilon^2) + \frac{B^2c^4}{G M^2} \epsilon^6 r^2 + o_2(r^2),\nonumber\\
  \mbox{sgn}(Q) \varphi(r)&=&\frac{3Mc^2}{2|Q|}   - \frac{A\epsilon^3}{2M} \frac{c^4}{G^{3/2}}r + O(r^3),
\end{eqnarray}
and near $r=\infty$,
\begin{eqnarray}
  \label{eq:asymptotics-near-infty}
  f(r)^2&=&1-\frac{2GM}{c^2r} + \frac{GQ^2}{c^4r^2}+O\left( \frac{1}{r^2} \right), \nonumber \\
  \varphi(r)&=& \frac{Q}{r} + O\left( \frac{1}{r^3} \right) .
\end{eqnarray}
Also,
\begin{align}
  (1-A\epsilon^2)& \leq  f(r)^2 < 1. &   \label{eq:e-xi-bnd}
\end{align}

The derivation of the Hamiltonian form of the Dirac equation for an electron on a spacetime 
with metric in the form \eqref{eq:background} and the separation 
under spherical symmetry is standard (\cite{thaller1992dirac} section 4.6, \cite{cohen1982general}). 
We define $x$ by 
\begin{equation}
  \frac{dx}{dr}=f(r)^{-2} ,  \ x(r=0) =0.
  \label{eq:x-To-r-Dir}
\end{equation}

Note that as $r\rightarrow \infty, x \rightarrow \infty$ and $\frac{x}{r} \rightarrow 1$. For, $\frac{x}{r} -1 = \frac{ \int_0^r  \frac{1}{f(s)^2} - 1 \ ds}{r}$. Given $\epsilon>0$, there is $r_1>0$ such that $|f(r)^{-2} -1|<\epsilon/2$ for all $r>r_1$. So, $|\frac{ \int_{r_1}^{r} (f(s))^{-2} -1 ds}{r}| \leq \frac{r-r_1}{r} \frac{\epsilon}{2}  \leq \frac{\epsilon}{2}$. Note that $f$ is bounded away from $0$. So, there is an $r_2>0$ such that for $r>r_2$, $\frac{\int_0^{r_1} (f(s)^{-2} -1 )ds }{r}< \epsilon/2$. So, for $r> \mbox{ max } \{r_1, r_2\}$, $|\frac{x}{r} -1 | < \frac{\epsilon}{2} + \frac{\epsilon}{2} = \epsilon$, by splitting the integration into $(0,r_1)$ and $(r_1,r)$.

From the previous paragraph it follows that $|\frac{1}{x} - \frac{1}{r} |  = \frac{1}{x} \frac{|x-r|}{r} \rightarrow 0 \cdot 1 = 0 $.
For later use, we prove the following lemma.
  
 \begin{lemma} For all $r>0$, $|\frac{f(r)}{r(x)} - \frac{1}{x}| \leq (1+f(0)^{-2}) \frac{1}{x}$. \end{lemma}

\begin{proof}
 Note that $x(r)= \int_0^r f(s)^{-2} ds$ and $f(0)<f(s)$ for all $s$. So, $f(s)^{-2}<f(0)^{-2}$. Hence, $x\leq r f(0)^{-2}$. So, $\frac{1}{r}\leq \frac{f(0)^{-2}}{x}$. Since $f(r) \leq 1$, it follows that $\frac{f(r)}{r} \leq \frac{f(0)^{-2}}{x}$. The lemma follows using triangle inequality.
 \end{proof}

The Dirac equation is \begin{equation} i\hbar \partial_t \psi = \mathcal{H}\psi , \end{equation} where  
the Dirac Hamiltonian $\mathcal{H}$ 
is a direct sum (indexed by $j,m_{j},k$) of operators $H$ (for ease of notation, we do not write the subscripts on $H$) 
\begingroup
  \renewcommand*{\arraystretch}{2}
  \begin{equation}
    H:= \begin{bmatrix}    mc^2 f(r) - e \varphi(r)  & -c
      \hbar \dfrac{d}{dx} + \hbar c f(r) \dfrac{\kappa}{r}   \\ \hbar c \dfrac{d}{dx} + \hbar c f(r) \dfrac{\kappa}{r}   &  - m c^2 f(r) - e \varphi(r) \end{bmatrix},
    \label{eq:Reduced-Dirac-Final}
  \end{equation}
\endgroup
each of which acts on $L^2((0,\infty);dx)^{2}$. This is equation (4.3) 
in \cite{cohen1982general}, with $f$ instead of $A$, $r$ instead of $S$ and with the constants $\hbar,c$ in the appropriate places.

The indices takes values with $j \in \left\{ \frac{1}{2}, \frac{3}{2}, \frac{5}{2}, \ldots \right \}$,
$ m_{j} \in \left\{ -j,-j+1,\ldots,j \right\}$ ,\\ $\kappa \in \left\{  -(j+1/2),(j+1/2) \right\}$. So, $\kappa$ takes non-zero integer values.

Letting $f=1$ and $\varphi=0$, we obtain $H_0$, the reduced free Dirac operator on Minkowski space.  Let $\sigma^1= \begin{bmatrix} 0 & 1 \\ 1 & 0 \end{bmatrix}, \sigma^3= \begin{bmatrix} 1 & 0 \\ 0 & -1 \end{bmatrix}$ be two of the Pauli matrices. Then, 
   $H=H_{0}+V$, where 
   \begin{equation}
     V:=\left(\frac{f(r)}{r(x)} -\frac{1}{x}\right)  \kappa \hbar c \sigma^1+ \left(-1+f(r)\right)mc^2 \sigma^3  -e\varphi(r) .
     \label{eq:Pot-Dir-Ess}
   \end{equation}

\section{Essential Self-adjointness}
\label{sec:esa}

  A symmetric operator with a dense domain in a Hilbert space is said to be 
  \emph{essentially self-adjoint} if the closure of the operator is self-adjoint;
  that is, the operator has one and only one self-adjoint extension.

We prove the following theorem. 
\begin{theorem}
  The reduced Dirac Hamiltonian $H$ is essentially self-adjoint on
the domain of compactly supported functions $D_c=C^\infty_c(0, \infty)^{2} $ in $ L^2((0,\infty);dx)^{2}  $.
  \label{thm:Dirac-esa-th}
\end{theorem}

First, we provide a definition, to align with Weidmann's book (\cite{weidmann1987spectral}). The $J$ there is the negative of our $J$ below. This
does not change the hypothesis on $P$ or the conclusions of the theorems we use, 
because we may apply those theorems to our $-\tau$.

\begin{defn}
  Suppose that for each $x$ in an interval $ (a,b)$,  $P(x)$ is a real symmetric matrix.
  By a \emph{Dirac-type differential expression} $\tau$ associated 
  to $P(x)$, we mean, for $u:(a,b) \rightarrow \C^2$,  
  \begin{equation}
  \tau u := Ju'+ Pu
    \label{eq:Tau-Dir-specific}
  \end{equation}
  \label{def:Dirac-type} where $J= \begin{bmatrix} 0 & -1 \\ 1 & 0 \end{bmatrix}$.
\end{defn}

A version of Weyl's alternative for Dirac-type systems is 
\begin{theorem*}[\cite{weidmann1987spectral}, Theorem 5.6 ]
  Suppose $\tau$ is a Dirac-type differential expression. Then, either:
  \begin{enumerate}[label=(\alph*)]
    \item for every $\lambda \in \C$ all solutions of $(\tau - \lambda ) u =0$ lie 
      in $L^2$ near $b$ (that is, for every solution $u$ there is a $c \in (a,b)$ 
      such that $u\in L^2 ((c,b),dx )$,
      or :
     \item for every $\lambda \in \C \setminus \R$ there exists a unique (up
       to a multiplicative constant) solution $u$ of $(\tau - \lambda )u=0$ which 
       is in $L^2$ near $b$. 
  \end{enumerate}

  In the first case, $\tau$ is said to be in the limit circle case (LCC)
  at $b$ and in the second case, $\tau$ is said to be in the limit point case (LPC).
  Similar result holds at the end-point $a$.
  \label{thm:Weyls-alternative}
\end{theorem*}

We now present a proof of theorem $\ref{thm:Dirac-esa-th} $.

\begin{proof} 

By [\cite{weidmann1987spectral}, Theorem 5.7],  $\tau$ has deficiency indices $(0,0)$ if 
it is in the limit point case at both $a$ and $b$. If the deficiency indices are $(0,0)$,
then the minimal operator (defined by $\tau$ 
by taking the closure of the operator $\tau$ defined originally 
on $C^{\infty}_{c}(0,\infty)^{2}$) is self-adjoint and therefore the only self-adjoint extension.
If $\tau$ is defined on $(a,\infty)$, then $\tau$ is in the LPC at $\infty$ (corollary of Theorem 6.8 \ in \cite{weidmann1987spectral}).
Let us determine the type at the other end point $0$. We will show that $(a)$ is not the case. For this, we take 
$\lambda=0$ and will show that not all solutions are in $L^2$ near $x=0$. We write 
\begin{equation}
  H = \hbar c \begin{bmatrix} 0&-  \ d_x +  f \frac{\kappa}{r}  \\ \ d_x + f \frac{\kappa}{r}  &  0 \end{bmatrix}  +  \begin{bmatrix} c^2 f m - e \varphi & 0 \\ 0 & -c^2 m f -  e \varphi  \end{bmatrix}.
  \label{eq:split-it}
\end{equation}
Since $|\varphi(r)|$ is decreasing 
and $|\varphi(0)|$ is finite and $f(r)$ is 
increasing with $\lim_{r\rightarrow \infty} f(r)=1$, the second matrix is a bounded operator. 
Essential self-adjointness is not affected by bounded perturbations, by the Kato-Rellich theorem. 
Therefore, in order to determine the type at $x=0$ we need only look at the first matrix. So, consider
\begin{equation}
  \begin{bmatrix} 0&-  \ d_x +  f(r) \frac{\kappa}{r}  \\  \ d_x + f(r) \frac{\kappa}{r}& 0  \end{bmatrix} \PM{  u_1(x) \\ u_2(x) } = 0.
  \label{eq:LPat0-1}
\end{equation}
This simplifies to two decoupled equations,
\begin{align}
  - u_2'(x) + f(r) \frac{\kappa}{r} u_2(x) &= 0, &
  u_1'(x) + f(r) \frac{\kappa}{r} u_1(x) &= 0.
  \label{eqa:LPat0compu}
\end{align}
Then,
\begin{align}
  u_2(x) &= c_2 e^{ \kappa \int_{x_0}^{x}  \frac{ f(r(s))}{r(s)}ds}    &
  u_1(x) &= c_1 e^{-\kappa \int_{x_0}^{x}  \frac{f(r(s))}{r(s)}ds}.
  \label{eq:LPat0compu-2}
\end{align}
where $x_0 \in (0,\infty)$ is fixed below and $c_1, c_2$ are any real numbers.

Using equation \eqref{eq:asymptotics-near-0}, 
$f(0)=\sqrt{1-A\epsilon^2}$, where $\epsilon$ is defined 
by equation \eqref{eq:Epsi}. Further, using the fact that $f(r)$ is 
increasing, and by continuity, given an $\eta>0$, 
there exists a $\delta>0$ such that, $\forall r \in (0,\delta)$, it follows that
\begin{equation}
  b_1 := \sqrt{1-A\epsilon^2}< f(r) < \sqrt{1-A\epsilon^2} + \eta =: b_2. 
  \label{eq:LPat0compu-3-temp}
\end{equation}
Let $x_0=x(r=\delta)$. Then,  $ \forall x \in (0, x_0)$, $b_2^{-2} < f(r)^{-2} < b_1^{-2} $.
Using \eqref{eq:x-To-r-Dir}, $b_2^{-2}r < x < b_1 ^{-2} r $.
So, $\dfrac{b_2^{-2}}{x} < \dfrac{1}{r} <\dfrac{ b_1^{-2} }{x} $. Using  \eqref{eq:LPat0compu-3-temp},
 $\dfrac{ b_1 b_2^{-2} }{x} < \dfrac{f(r)}{r} < \dfrac{b_2 b_1^{-2}}{x}$. So, 
 $   b_1 b_2^{-2} \ln( \frac{x}{x_0}  ) \leq \int_{x_0}^{x} \frac{ f(r(s)) }{r(s)} ds \leq b_2 b_1^{-2}\ln(\frac{x}{x_{0}}) $.

Recall that $\kappa$ is a non-zero integer. Suppose $\kappa\geq 1$. Then, multiplying the previous equation with $-\kappa$, which is negative and using the definition of $u_1$ we get
$|u_1(x)| \geq  |c_1| e^{-b_2 b_1^{-2} \kappa \ln(\frac{x}{x_{0}})}$. 
So, we obtain $|u_1(x)| \geq |c_1| (\frac{x}{x_0})^{-\kappa b_2 b_1^{-2} }  $.

Similarly, if $\kappa \leq -1$, we multiply by $\kappa$, which is negative, and obtain $|u_{2}(x)| \geq  |c_2| (\frac{x}{x_0})^{\kappa b_2 b_{1}^{-2}}. $ 

Note that, $b_2 b_1^{-2} = \frac{b_2}{b_1} \frac{1}{b_1} \geq \frac{1}{b_1} >  1$, where we have used $b_2 > b_1 $  and $0< b_1 < 1$.
Since $|\kappa| \geq 1$, we have thus shown that because of $u_1$ in the case $\kappa \geq 1$ and $u_2$ in the case $\kappa  \leq -1$, for every $\kappa$ there is a particular solution that is not in  $L^2( (0,x_0),dx) ^{2}$. 

Thus, we have verified that $H$ is in the limit point case at 
the boundary point $0$. So $H$ is in LPC at both boundary points $0,\infty$,
and therefore $H$ is essentially self-adjoint on $C^{\infty}_{c}(0,\infty)^{2}$. 

This ends the proof of Theorem \ref{thm:Dirac-esa-th}.

\end{proof}

\section{Spectrum}
\label{sec:spec}

\subsection{Essential spectrum}
\label{ssec:essspec}
The \emph{essential spectrum} of a self-adjoint operator consists of accumulation points of the spectrum and the 
eigenvalues whose eigenspace is infinite dimensional.
The spectrum of the free Dirac operator on Minkowski space is $(-\infty,-mc^2] \cup [ mc^2, \infty)$. So this set is also the essential spectrum.  

\begin{theorem}
  The essential spectrum of $H$ is $(-\infty,-  mc^2] \cup [ mc^2,\infty)$. 
  \label{thm:Ess-Spec}
\end{theorem}

Thaller \cite{thaller1992dirac} (section 4.3.4) provides theorems showing invariance of essential spectrum, but of the full operator. It is possible to adapt that to our situation, which concerns the radial part. Theorems in Weidmann's book \cite{weidmann1987spectral} (chapter 16) are better suited as they deal with the reduced operators.

  \begin{theorem*}[Theorem 16.5, \cite{weidmann1987spectral}]
  Assume $\tau$ is a Dirac-type differential equation and that $\tau$ is regular at $a$ and that $b= \infty$. If $P(x) \rightarrow P_0$ for $x \rightarrow \infty$ and $\mu_{-} \leq \mu_{+}$ are the eigenvalues of $P_0$, then for every self-adjoint realization $A$ of $\tau$, we have, (with $\sigma_e(A)$ the essential spectrum), $$ \sigma_e(A) \cap (\mu_{-}, \mu_{+})  = \phi.$$
   \label{thm:Weid-essspec1}
  \end{theorem*}
  
  Our $\tau$ is not regular at $x=0$ because although $x=0$ is finite, the entries of $P(x)$ have $1/x$ terms and so are 
  not locally integrable on $[0, \infty)$. The definition of 'regular' is in the end of introduction of chapter 3 and in chapter 2 of the book.
  
  The theorem is still useful because of the \emph{decomposition method}. This is used in Korollar 6.13 of 
  \cite{weidmann1971oszillationsmethoden}.
  In the remark after theorem 11.5 in \cite{weidmann1987spectral}, it is noted:
  $\sigma_e(A) = \sigma_e(A_a) \cup \sigma_e(A_b)$ where $A, A_a, A_b$ are arbitrary self-adjoint extension of $T_0, T_{a, c,0}, T_{c, b, 0}$ which are the minimal operators on $(a,b)$ (the original interval), $(a, c), (c,b)$ respectively, 
  with $c \in (a,b)$.
  
  So, it is enough to show that $A_a$ has no essential spectrum. Now, $\tau$ on $(a, c)$ has one endpoint singular and other regular. So, by theorem $16.4$ (we have not reproduced it here) and the observation after the proof of the theorem, it is enough to show that the Pr\"ufer variable $\theta(x, \lambda)$ 
  (defined using $u = \rho(x) \Theta(x) \mbox{ where } \Theta(x) = \begin{bmatrix} \cos(\theta(x)) \\ \sin(\theta(x)) \end{bmatrix}$) remains bounded as $x\rightarrow 0=a$ for all spectral parameters $\lambda.$ 
  
 According to the introduction to chapter $16$ of \cite{weidmann1987spectral}, $\theta' = ( G(x) \Theta(x), \Theta(x)) , \mbox{ with }  G(x) = \frac{1}{2} (\lambda - P(x)) $ in our situation.   
 So, $\theta' = \cos^2(\theta) (\lambda - P_{11}) + 2 \cos(\theta) \sin(\theta) P_{12} + \sin^2(\theta) (\lambda - P_{22}) =: R$.
 In order to argue that $\theta(x) $ is bounded for $x \in (0, 1)$, we will convert this into a dynamical system on a cylinder, similar to what is done in \cite{kz2014dpezgkn} section VII.  
 
 As seen below, for us $P_{11}, P_{22}$ are bounded for $x \in (0, \infty)$ and $x P_{12}$ is bounded on $(0, c)$ for $c< \infty.$ 
 Let $s$ be a variable change of $x$ such that $\tan(s) = x$.
 So, $s\in (0, \pi/4).$ 
 Now, $s\theta' = s R$. We want this to be our first equation: $\frac{d\theta}{dm} = sR$. This in turn implies by chain rule that $\frac{dx}{dm} = s$ and therefore $sec^2(s) \frac{ds}{dm} = s$ and so, $\frac{ds}{dm} = s \cos^2(s).$ 
 
 Thus we have a dynamical system in dependent variables $\theta, s$ and independent variable $m$. Since $R$ dependence on $\theta$ is only through $\cos(\theta), \sin(\theta)$, this dynamical system can be considered to be on the cylinder $s\in [0, \pi/4] , \theta \in [0, 2\pi] $. 
 
 We see that $sR$ can be continued into $s=0$. So, right hand side of the dynamical system is at least continuously defined on the compact cylinder. Further, on the edge $s=0$, orbits lie on the edge, as $\frac{ds}{dm}=0$. As $m$ decreases, the orbits must move towards $s=0$ as $ds/dm >0$. Further, they cannot keep on winding around the cylinder but they must have a approach critical point on the left edge. This shows that $\theta$ stays bounded as $s \rightarrow 0$, that is, $x \rightarrow 0$. 

We will also use the following theorem.

  \begin{theorem*}[Theorem 16.6, \cite{weidmann1987spectral}]
  Assume $\tau$ is a Dirac-type differential equation and and that $b= \infty$.  Assume that for some $d \in (a,\infty)$,  $$x^{-1} \int_{d}^{x} |P(t) - P_0| dt \rightarrow 0 \ \mbox{ for } x \rightarrow \infty $$  and $\mu_{-} \leq \mu_{+}$ are the eigenvalues of $P_0$, then for every self-adjoint realization $A$ of $\tau$, we have, (with $\sigma_e(A)$ the essential spectrum), $$ (\mu_{-}, \mu_{+})^{c}  \subset \sigma_e(A).$$
   \label{thm:Weid-essspec2}
  \end{theorem*}
\begin{proof}(of theorem (\ref{thm:Ess-Spec}))

We start by observing that $\tau = (\hbar c)^{-1} H$ on $(a, b) = ( 0, \infty)$ is a Dirac-type differential expression with 
$P(x) = \begin{bmatrix} \frac{mc}{\hbar} f(r) - \frac{e}{\hbar c} \varphi(r)  &  f(r) \frac{ \kappa} {r} \\
f(r) \frac{ \kappa}{r} &  - \frac{mc}{\hbar} f(r) - \frac{e}{\hbar c} \varphi(r) \end{bmatrix}$.

Let $P_0 = \begin{bmatrix} \frac{mc}{\hbar} & 0 \\ 
                                0  &  - \frac{mc}{\hbar} \end{bmatrix}.$
So, $\mu_{\pm} = \pm\frac{mc}{\hbar}$ are the eigenvalues of $P_0$. Further, as $x \rightarrow \infty$, $r \rightarrow \infty$ and so $f \rightarrow 1, \varphi \rightarrow 0$ and so $P(x) \rightarrow P_0$. 

Let $d=1.$ Then $ \int_{d}^{ x} f(r(t)) \frac{\kappa}{r} dt =  \int_{r(d)}^ {r(x)} f(\tilde{t}) \frac{ \kappa}{\tilde{t}} f(t)^{-2} d \tilde{t} $. Since $f$ is bounded, this is bounded by $C \kappa ( \ln(\frac{r(x)}{r(d)})$, were $C(f)$ is a constant independent of $x$. We noted earlier that $\frac{x}{r} \rightarrow 1$ as $r \rightarrow \infty$. So, $\ln(x) - \ln(r) \rightarrow 0 $. So, $C \kappa \frac{1}{x} \frac{ \ln(r(x)}{r(d)} \rightarrow  C \kappa  \frac{1}{x} \frac{\ln(x)}{\ln(r(d))} \rightarrow 0$.

Note that $\varphi(r) = \frac{Q}{r} + O(\frac{1}{r^3})$ as $r \rightarrow \infty$.
 Now, $\int_{d}^{x} |\varphi(r(t))| dt $ is at most $ \int_{r(d)}^{r(x)} |\varphi(r)| f(\tilde{t})^{-2} d \tilde{t}$ which is at most $ C (|Q| \ln( \frac{\ln(r(x)}{r(d)}) + M (\frac{1}{r^2}\big|_{r(d)}^{r(x)}) ) $, for sufficiently large $d$. Here, $C(f), M(\varphi)$ are constants. So, $\frac{1}{x} \int_d^x |\varphi(r(t))| dt \rightarrow 0$.
Finally, from the asymptotics for $f^2$, we see $f(r) = 1 + O(1/r)$ as $r \rightarrow \infty.$  
So, $\int_d^x |f(r(t)) - 1| dt = \int_d^x |f(\tilde{t}) - 1| f(\tilde{t})^{-2} d\tilde{t}  \leq M_2  \frac{ \ln(r(x))} {\ln(r(d))} $ where $M_2(f)$. 
As before, $\frac{1}{x} \int_d^x |f(r(t)) - 1| dt  \rightarrow 0$.
All in all, we have shown that $\frac{1}{x} \int_d^x |P(t) - P_0| dt \rightarrow 0$, for a sufficiently large $d$.
 
So, the latter theorem allows us to conclude that $ ( -\frac{mc}{\hbar}, \frac{mc}{\hbar} )^{c} \subset \sigma_e( (\hbar c)^{-1} H) $. And, the former theorem allows us to conclude (via the decomposition method argument present above),  
 $ ( -\frac{mc}{\hbar}, \frac{mc}{\hbar} ) \cap \sigma_e( (\hbar c)^{-1} H) = \phi $.  Thus, theorem \ref{thm:Ess-Spec} is proven.

\end{proof}

\subsection{Continuous Spectrum}
\label{ssec:contspec}
The spectral measure associated with a self-adjoint operator can be written as a sum of 
pure-point, absolutely continuous and singular measures. This results in the decomposition of the Hilbert space. The absolutely continuous spectrum of the operator is defined to be the spectrum of the 
the operator that is its restriction to the absolutely continuous part of the Hilbert space.

For the free Dirac operator, the unitary transformation to a multiplication operator is 
known using Fourier Transform. That helps one determine the spectrum and in particular, on $(-\infty,-mc^2) \cup (mc^2, \infty)$ it is shown to be 
purely absolutely continuous. See theorem 1.1 \ in \cite{thaller1992dirac}.
In this section, we show that this is true even for our case. 

\begin{theorem}
The reduced Dirac Hamiltonian $H$ has 
  purely absolutely continuous spectrum in $(-\infty,-mc^2) \cup (mc^2,\infty)$.
  \label{thm:PAC-DIR}
\end{theorem}

\begin{proof}
  We prove this theorem using a result from Weidmann's book, which 
  we have paraphrased below for our form of the Dirac type operator, which is negative of what is there.

  \begin{theorem*}[Theorem 16.7, \cite{weidmann1987spectral}]
    Consider a Dirac-type expression $\tau$ (definition \ref{def:Dirac-type}) on $(a,\infty)$,
    for which the matrix $P(r)$ can be written as $P_1(r)+P_2(r)$, where for some $c\in (a,\infty)$
    the components of $P_1(r)$ are in $L^1( (c,\infty))$,  and the components of $P_2(r)$ are of bounded
    variation in $[c,\infty)$. Suppose also that 
      \begin{equation}
	\lim_{r\rightarrow \infty} P_2(r) = \begin{bmatrix} \mu_{+} &0\\0& \mu_{-} \end{bmatrix},  \mu_{-} \leq \mu_{+}.
	\label{eq:lim-P2}
      \end{equation}
   Then, every self-adjoint realization of $\tau$ has purely absolutely continuous spectrum
   in $(-\infty, \mu_{-}) \cup (\mu_{+},\infty)$.
    \label{thm:Weid-absc}
  \end{theorem*}

 In our case, consider $\tau $ as earlier (ie, $(\hbar c)^{-1} H)$ and let $P_1 =0 , P_2(x) = P(x).$
We have shown in the proof of theorem $(\ref{thm:EValClustering})$ that $(f(r)/r)' = O(1/x^2)$, thus allows us to conclude $(f(r)/r)' $ is in $L^1(c, \infty)$. Using the definition, $\varphi'(r)  =  Q \zeta_{\beta}'(Q^2/(2r^4)) \frac{1}{r^2}$. Note that by the properties required for $\zeta$, $\zeta' \leq \zeta(\mu)/\mu $, so it is bounded near $\mu =0$. Hence, $\varphi'(r) = O(\frac{1}{r^2})=O(\frac{1}{x^2}).$
Now, $f(r)^2 = 1 - \frac{2G}{rc^4}  \int_0^{r} \zeta_{\beta}(Q^2/(2s^4)) s^2 ds $. The integral term (called mass function) is shown in \cite{stz2011} to be bounded. So, bearing in mind that $f(r)$ is bounded, to show $f'(r) =O(1/x^2)$ all we need is to show that the derivative of the integral term is $O(1/r)$. Indeed, it is. In fact, $\zeta_{\beta} ( Q^2/(2r^4)) r^2 = O(1/r^4) r^2 = O(1/r^2) $. 

Thus functions $f(r)$, $\varphi(r)$, $f(r)/r$, considered as
  functions of $x$, are of bounded variation in $[1,\infty)$. And so, $P_2$ satisfies the required properties and $\mu_{-}=-\frac{mc^2}{\hbar c}, \mu_{+} = \frac{mc^2}{\hbar c}$ using the asymptotics at $x= \infty.$ 
  Thus the theorem is proven.
  
\end{proof}

\subsection{Existence and number of Eigenvalues}
\label{ssec:discspec}
 
In \cite{HMRS}, the authors built on the earlier ideas of Birman and Kurbenin,
to relate the finiteness of the spectrum in the gap of the essential spectrum of a Dirac operator to the oscillatory properties
of an associated second order ordinary differential equation (ODE). This ODE involves a positive-linear
functional, which by definition, is a linear functional on space of square matrices that evaluates to a positive
number on matrices that are both symmetric and positive definite. Such functionals
are finite sums of functionals of the form $g(B)=(Bu,u)$, for some non-zero vector $u$.

In this section we prove the following theorem:
\begin{theorem}
 The gap spectra $\sigma(H) \cap (-mc^2, mc^2) $ is non-empty and infinite.
  \label{thm:EValExistence}
\end{theorem}
We remark that since the spectrum in the gap $(-mc^2, mc^2)$ consists only of eigenvalues, this 
theorem shows that there are an infinite number of eigenvalues.

Let $J=\begin{bmatrix}0&-1\\1&0\end{bmatrix}$ and $P(x) =\begin{bmatrix} V_2(x) -c_2 & p(x) \\p(x) & V_1(x) + c_1
 \end{bmatrix}$, where $c_1,c_2$ are positive numbers, and $V_1(x),V_2(x),p(x)$ are real-valued, 
locally integrable functions defined on $(0,\infty)$. 
Let $L_1$ be any self-adjoint extension of the minimal operator defined by the $
  Jy' - P(x)y$ for compactly supported smooth $y$.

\begin{theorem*} [\cite{HMRS}, 2.3]
      Suppose that $d>0$ and $g$ is a positive linear functional on  real $n\times n$
matrices and assume $P$ is locally absolutely continuous. Then, $\sigma(L_1) \cap (-d,d)$ is infinite if the scalar
differential equation 
\begin{equation}
  -g[I]z'' + g\left[P^2 - d^2 I^2 + \frac{P'J-JP'}{2}\right]z=0.
  \label{eq:gapOde}
\end{equation}is oscillatory at zero or infinity.
\end{theorem*}

\begin{proof}

We will apply the theorem to $L_1=(\hbar c)^{-1} H$. We proceed as in equations (2.6), (2.7) and Example 1 in \cite{HMRS}.
Let us take $g[B]=(Bu,u)$ with the vector $u=(1,0)^{T}$. So, $g[I]=1$, $g[B]=B_{11}$.

 In our case, \begin{equation}
   \Gamma(x):= g\left[P^2-d^2I+\frac{P'J-JP'}{2}\right]=(-mc^2+e\varphi)^2 (\hbar c)^{-2} +\left(\frac{kf}{r}\right)^2 - d^2 + k\left(\frac{f}{r}\right)'.
  \label{eq:gapOde2}
\end{equation}
Note that as before a prime indicates differentiation in the variable $x$.
From Corollary $37$ in \cite{dunford1963linear}, every solution of $-z''+\Gamma(x) z=0$
has an infinite number of zeros on a neighbourhood $[a,\infty)$ of $\infty$ if 
  $\lim_{x\rightarrow \infty} x^2 \Gamma(x) < -\frac{1}{4}$. That is, the differential equation
  is oscillatory at $\infty$.
  
  As  $x \rightarrow \infty $, we have, $ \varphi \sim \frac{Q}{r}, f \sim 1, \frac{1}{r} \sim \frac{1}{x} $ and $(\frac{f}{r})' = O(1/x^2)$ (which we showed in the previous subsection). Take $d= mc^2 (\hbar c)^{-1}$. This will cancel the $(mc^2 )^2$ term.  So we see that only non-finite contribution to the limit is   $ \lim_{x\rightarrow \infty} x^2 (-2me\varphi(r)\hbar^{-2})=-\infty,$
  if $-eQ<0$.
  And so, $\lim_{x\rightarrow \infty} x^2 \Gamma(x) = - \infty$.

  If $-eQ>0$, we may take $g[B]=(Bu,u),u=(0,1)^{T}$. Then, the only changes in 
  the expression for $\Gamma(x)$ is that $-mc^2$ becomes $+mc^2$ 
  and $+\kappa (f/r)'$ turns into $-\kappa (f/r)'$. So, the same argument above helps us
  conclude that the resulting differential equation is oscillatory.
  
  Therefore, there are infinitely many eigenvalues of $\hbar^{-1}c^{-1} H$ 
  in $(-mc^2 (\hbar c)^{-1}, mc^2 (\hbar c)^{-1}) $, which proves our theorem.  
\end{proof}

We note that similar ideas are seen in the appendix C of \cite{belgiorno2000nakedanom}.

\subsection{Clustering of eigenvalues}
\label{ssec:clustering}
In this section we prove the following theorem:
\begin{theorem}
 Assume the spacetime charge $Q>0$. Then, in $\sigma(H)$,  $mc^2$ is a cluster point and $-mc^2$ is not a cluster point.
  \label{thm:EValClustering}
\end{theorem}

The following theorem is in the notation we mentioned in the previous subsection for defining $L_1$. It is paraphrased to fit our application. In particular, in the original theorem, (ii) has $x^2V(x) \rightarrow -\infty$ and concludes $-\tilde{c}$ is a cluster point and $\tilde{c}$ is not. The comment made after equation (4.3) there asserts the validity of our paraphrasing. 

The theorem assumes that the essential spectrum of $L_1$ is $(-\infty, -\tilde{c}] \cup [\tilde{c}, \infty)$ and that there are infinitely many eigenvalues in the gap $[-\tilde{c},\tilde{c}]$. 

Note that $\tilde{c}>0$ is a positive number. The original source uses $c$ instead of $\tilde{c}$.

\begin{theorem*} [\cite{HMRS}, 4.1]
      Suppose that $V_1=V_2=V$, $c_1=c_2=\tilde{c}$, $p$ is locally absolutely continuous and 
      \begin{enumerate}[label=(\roman*)]
      \item $p(x)=O(1/x), p'(x) = O(1/x^2)$ as $x \rightarrow \infty,$
      \item $V(x) \rightarrow 0$ and $x^2 V(x) \rightarrow \infty$ as $x \rightarrow \infty.$
      \item 
      In a neighbourhood of $0$, $p(x) \geq |V(x)| + \tilde{c} $ or $p(x) \leq - \tilde{c} - |V(x)|$. 
      \end{enumerate}
      Then $\tilde{c}$ is a cluster point of eigenvalues and $-\tilde{c}$ is not.
\end{theorem*}

\begin{proof} (Of theorem \ref{thm:EValClustering}) \label{pf:clustering}
As before we apply the theorem above to $L_1  = (\hbar c)^{-1} H.$ Then, $V(x) = \frac{e \varphi}{\hbar c} + \frac{mc^2}{\hbar c } (1-f)$ , $p(x) = -\frac{kf}{r}$, $\tilde{c}= \frac{mc^2}{\hbar c}$. The $1$ in $V$ is from having to split up to fit the notation of the theorem.
It follows from the near $\infty$ asymptotics we gave for $f$ that $f$ approaches $1$ and by the observation after equation \eqref{eq:x-To-r-Dir} that $\frac{1}{r} \sim \frac{1}{x}$ as $r \rightarrow \infty$ that $p(x) = O(1/x)$ as $x \rightarrow \infty$. 
Now, using \eqref{eq:x-To-r-Dir}, $p'(x) = kf^3/r^2 -\frac{kf(r)}{2r} \frac{d (f(r))^2}{dr}$.
Using equation $ \eqref{eq:xiandm}$, 
$\frac{d(f^2)}{dr} = \frac{-2G}{c^4} \left( \frac{1}{r^2} \int_0^r \zeta_{\beta} (w(s)) s^2 ds + \frac{-1}{r}\zeta_{\beta} (w(r)) r^2 \right), $ 
where $w(r) = \frac{Q^2}{2r^4}$. Using $\zeta(\mu) \leq \mu$, we see that the integral term is bounded and that the second term is bounded by $\frac{1}{r} \frac{1}{\beta^4} \beta^4 w r^2.$ which because of the $w$ is $O(1/r^3)$. So, overall $p'(x) = O(1/x^2)$ as $x \rightarrow \infty$. We have thus established $(i)$.

That $V(x) \rightarrow 0$ is evident from the asymptotics of $f^2, \varphi$ near $\infty$. Also, $x^2 e \varphi/{\hbar c} \rightarrow + \infty$ as $\varphi >0$ as $Q >0$ and $\varphi(r) \sim Q/r$ near $\infty$. Using the asymptotics for $f^2$, we see that  $(1-f) = \frac{GM}{c^2} \frac{1}{r} + O(1/r^2)$ near $\infty$. So, $x^2 (1-f)  \rightarrow \infty.$
Thus $(ii)$ is established,

Finally, $(iii)$ is true because $p$ blows up like $\frac{1}{r}$ near  $r=x=0$ while $V$ is a uniformly bounded function on $(0, \infty)$. 

Thus we get that $-\tilde{c} = - \frac{mc^2}{\hbar c}$ is not a cluster point and $\tilde{c} = \frac{mc^2}{\hbar c}$ is a cluster point of eigenvalues of $(\hbar c)^{-1} H$. The theorem follows from this.

\end{proof}

\section{Summary}
\label{sec:summary}
We proved essential self-adjointness, found the essential and continuous spectrum and showed the existence of infinitely many eigenvalues and their clustering for the Dirac-Hamiltonian on a spacetime with mild singularity, of which an example is the Hoffmann spacetime coming from the Born-Infeld Lagrangian. A work in progress is the numerical estimation of the eigenvalues. Also of interest would be a non-test particle approach.

\bibliographystyle{plain}
\bibliography{refs-paperSubmission}

\end{document}